\documentclass[10pt,journal]{IEEEtran}
\usepackage{ifpdf}
\ifpdf
	\usepackage[pdftex]{graphicx}
    \usepackage[update]{epstopdf}
\else
	\usepackage{graphicx}
\fi

\usepackage{amsmath}
\usepackage{amsfonts}
\usepackage{amssymb}
\usepackage{amstext}
\usepackage{latexsym}
\usepackage{color}
\usepackage{cite,setspace}
\usepackage{multirow}
\usepackage{breqn,url}
\usepackage{flushend}

\newtheorem{theorem}{\textbf{Theorem}}
\newtheorem{prop}{\textbf{Proposition}}
 
\newtheorem{definition}{\textbf{Definition}}%[section]

\renewcommand{\vec}[1]{\mbox{\boldmath$#1$}}

\allowdisplaybreaks[1]

\begin{document}

\title{Characterizing the Rate Region of the $(4,3,3)$ Exact-Repair Regenerating Codes}

\author{Chao Tian,~\IEEEmembership{Senior Member,~IEEE}\thanks{Chao Tian is with AT\&T Labs-Research, Florham Park, NJ 07932. }
}
\maketitle

\begin{abstract}
Exact-repair regenerating codes are considered for the case $(n,k,d)=(4,3,3)$, for which a complete characterization of the rate region is provided. This characterization answers in the affirmative the open question whether there exists a non-vanishing gap between the optimal bandwidth-storage tradeoff of the functional-repair regenerating codes ({\em i.e.,} the cut-set bound) and that of the exact-repair regenerating codes.  To obtain an explicit information theoretic converse, a computer-aided proof (CAP) approach based on primal and dual relation is developed. This CAP approach extends Yeung's linear programming (LP) method, which was previously only used on information theoretic problems with a few random variables due to the exponential growth of the number of variables in the corresponding LP problem. The symmetry in the exact-repair regenerating code problem allows an effective reduction of the number of variables, and together with several other problem-specific reductions, the LP problem is reduced to a manageable scale. For the achievability, only one non-trivial corner point of the rate region needs to be addressed in this case, for which an explicit binary code construction is given. 
\end{abstract}

\section{Introduction}

Erasure codes can be used in data storage systems that encode and disperse information to multiple storage nodes in the network (or multiple disks inside a large data center), such that a user can retrieve it by accessing only a subset of them. This kind of systems is able to provide superior availability and durability in the event of disk corruption or network congestion, at a fraction of the cost of the current state of art storage systems based on simple data replication. When data is coded by an erasure code, data repair becomes more involved, because the information stored at a given node may not be directly available from any one of the remaining storage nodes. One key issue that affects the overall system performance is the total amount of information that the remaining nodes needs to transmit to the new node.

Dimakis {\em et al.} \cite{Dimakis:10} proposed the framework of regenerating codes to address the tradeoff between the storage and repair bandwidth in erasure-coded distributed storage systems. In this framework, the overall system consists of $n$ storage nodes situated in different network locations, each with $\alpha$ units of data, and the content is coded in such a way that by accessing any $k$ of these $n$ storage nodes,  the full data content of $B$ units can be completely recovered. When a node fails, a new node may access any $d$ remaining nodes for $\beta$ units of data each, in order to regenerate a new data node. %Note that any $d$ such node have to be able to serve as the data provider in the regenerating process, which places a more stringent but also more robust regenerating condition than requiring the existence of at least one of such $d$-node combinations. 

The main result in \cite{Dimakis:10} is for the so-called functional-repair case, where the regenerating process does not need to exactly replicate the original data stored on the failed node, but only needs to guarantee that the regenerated node can serve the same purpose as the lost node, {\em i.e.}, data reconstruction using any $k$ nodes, and being able to help regenerate new data nodes to replace subsequently failed nodes. It was shown that this problem can be cleverly converted to a network multicast problem, and the celebrated result on network coding \cite{Yeung:00} can be applied directly to provide a complete characterization of the optimal bandwidth-storage tradeoff. Furthermore, linear network codes \cite{Yeung:03} are sufficient to achieve this optimal performance.

The decoding and repair rules for functional-repair regenerating codes may evolve as nodes are repaired, which increases the overhead of the system. Moreover, functional-repair does not guarantee systematic format storage, which is an important requirement in practice. For these reasons, exact-repair regenerating codes have received considerable attention recently \cite{RashmiShah:11,RashmiShah:12:1,RashmiShah:12:2,Cadambe:11}, where the regenerated data need to be exactly the same as that stored in the failed node. 

The optimal bandwith-storage tradeoff for the functional-repair case can clearly serve as an outer bound for the exact-repair case. There also exist code constructions for the two extreme cases, {\em i.e.}, the minimum storage regenerating (MSR) point \cite{RashmiShah:11,RashmiShah:12:2,Cadambe:11}, or the minimum bandwidth regenerating (MBR) point \cite{RashmiShah:11,RashmiShah:12:1}, and the aforementioned outer bound is in fact achievable at these two extreme points. The achievability of these two extreme points immediately implies that for the cases $k\leq2$, the functional-repair outer bound is tight for the exact repair case. Also relevant is the fact that symbol extensions are necessary for linear codes to achieve the MSR point for some parameter range \cite{RashmiShah:12:2}, however  the MSR point can indeed be asymptotically (in $B$) achieved by linear codes for all the parameter range \cite{Cadambe:11}. It was also shown in \cite{RashmiShah:12:1} that when $k>2$, other than the two extreme points and a segment close to the MSR point, the majority the functional repair outer bound is in fact not strictly achievable by exact-repair regenerating codes. 

The non-achievability result reported in \cite{RashmiShah:12:1} was proved by contradiction, {\em i.e.},  a contradiction will occur if one supposes that an exact-repair code operates {\em strictly} on the optimal functional-repair tradeoff curve. However, it is not clear whether this contradiction is caused by the functional-repair outer bound being only asymptotically achievable, or caused by the existence of a non-vanishing gap between the optimal tradeoff of exact-repair codes and the functional-repair outer bound. In fact, the necessity of symbol extension proved in \cite{RashmiShah:12:2} and the asymptotically optimal construction given in \cite{Cadambe:11} may be interpreted as suggesting that the former is true. 

In this work, we focus on the simplest case of exact-repair regenerating codes, {\em i.e.}, when $(n,k,d)=(4,3,3)$, for which the rate region has not been completely characterized previously. A complete characterization of the rate region is provided for this case, which shows that indeed there exists a non-vanishing gap between the optimal tradeoff of the exact-repair codes and that of the functional-repair codes. The achievability part of this result shows that there exist exact-repair regenerating codes that are better than simply time-sharing between the MSR point and the MBR point.

As in many open information theoretical problems, the difficulty lies in finding good outer bounds, particularly in this problem with a large number of regenerating and reconstruction requirements. We rely on a computer-aided proof (CAP) approach and take advantage of the symmetry and other problem-specific structure to reduce the number of variables in the optimization problem. This approach builds upon Yeung\rq{}s linear programming (LP) framework \cite{Yeung:book}. As of our knowledge, this is the first time that the LP framework is meaningfully applied to a non-trivial engineering problem, which leads to a complete solution. More importantly, instead of only machine-proving whether an information theoretic bound is true or not as in \cite{Yeung:book}, we further develop a secondary optimization procedure to find an \textit{explicit information theoretic proof}. By solving the primary LP optimization problem, the tradeoff curve between the storage and repair bandwidth can be traced out numerically, which leads to the hypotheses of the bounding planes for the rate region. A secondary optimization procedure, which essentially solves the dual problem for these candidate bounding planes, directly yields an explicit information theoretic proof. Due to the duality structure in the LP problem, the optimization criterion in the secondary optimization problem can be selected arbitrarily, thus we can choose one that leads to the solution that we most desire. For this purpose, the $\ell_1$ norm is chosen that approximates the solution under the $\ell_0$ norm, the latter of which gives the sparsest solution and translates roughly to a converse proof with the least number of steps.

The rest of the paper is organized as follows. In Section \ref{sec:definition}, we provide a formal definition of the problem and review briefly the functional-repair outer bound.  The characterization of the rater region is given in Section \ref{sec:main}, together with the forward and converse proof. Section \ref{sec:CAP} provides details on the computed-aid proof approach, and Section \ref{sec:conclusion} concludes the paper.

\section{Problem Definition}
\label{sec:definition}

In this section we first give a formal definition of the regenerating code problem for the case $(n,k,d)=(4,3,3)$, and then introduce some notation useful for the converse proof. Somewhat surprisingly, we were not able to find such a formal definition in the existing literature, and thus believe it is beneficial to include one here (which can be generalized to other parameters).  The functional-repair outer bound is briefly reviewed and specialized to the case under consideration.

\subsection{Exact-Repair Regenerating Codes}
A $(4,3,3)$ exact-repair regenerating code is formally defined as follows, where the notation $I_n$ is used to denote the set $\{1,2,\ldots,n\}$, and $|A|$ is used to denote the cardinality of a set $A$.

\begin{definition}
\label{def:NKKcode}
An $(N,K_d,K)$ exact-repair regenerating code for the $(4,3,3)$ case consists of $4$ encoding functions $f^E_i(\cdot)$,  $4$ decoding functions $f^D_{A}(\cdot,\cdot,\cdot)$, $12$ repair encoding functions $F^{E}_{i,j}(\cdot)$,  and $4$ repair decoding functions $F^{D}_{i}(\cdot,\cdot,\cdot)$, where
\begin{align*}
f^E_i:I_N\rightarrow I_{K_d},\quad i\in I_4,
\end{align*}
each of which maps the message $m\in I_N$ to one piece of coded information, 
\begin{align*}
f^D_{A}:I_{K_d}\times I_{K_d}\times I_{K_d}\rightarrow I_N,\quad A\subset {I}_4\quad \mbox{and}\quad |A|=3,
\end{align*}
each of which maps $3$ pieces of coded information stored on a set $A$ of nodes to the original message, 
\begin{align*}
F^{E}_{i,j}:I_{K_d}\rightarrow I_{K},\quad j\in I_4,\quad\mbox{and}\quad i\in I_4\setminus \{j\},
\end{align*}
each of which maps a piece of coded information at node $i$ to an index that will be made available to reconstruct the data at node $j$, and
\begin{align*}
F^{D}_{j}:{I}_{K}\times{I}_{K}\times{I}_{K} \rightarrow {I}_{K_d},\quad \quad j\in{I}_4,
\end{align*}
each of which maps 3 such indices from the helper nodes to reconstruct the information stored at the failed node.  The functions must satisfy the data reconstruction conditions%\footnote{This condition enforces zero-error reconstruction. The asymptotic zero-error-probability condition may be relevant in practice, for which the main result in this paper also holds.}
\begin{align}
f_{{A}}^D\left(\prod_{i\in{A}}f^E_i(m)\right)=m,\quad m\in{I}_N,\,\,{A}\subset {I}_4\,\, \mbox{and}\,\, |{A}|=3,
\label{eqn:reconstructionzeroerror}
\end{align}
and the repair conditions
\begin{align}
F^D_{j}\left(\prod_{i\in{I_n\setminus \{j\}}}F^E_{i,j}\left(f^E_i(m)\right)\right)=f^E_j(m),\,\, m\in{I}_N,\quad j\in{I}_4.
\end{align}
\end{definition}

\vspace{0.2cm}
In the above definition, $N$ is the cardinality of the message set, and $\log N$ is essentially $B$. Similarly $\log K_d$ is essentially $\alpha$ and $\log K$ is $\beta$. To include the case when the storage-bandwidth tradeoff may be approached asymptotically, {\em e.g.}, the codes considered in \cite{Cadambe:11}, the following definition which utilized a normalized version of $\alpha$ and $\beta$ is further introduced. 

\begin{definition}
A normalized bandwidth-storage pair $(\bar{\alpha},\bar{\beta})$ is said to be $(4,3,3)$ exact-repair achievable if for any $\epsilon>0$ there exists an $(N,K_d,K)$ exact-repair regenerating code such that
\begin{align}
\bar{\alpha}+\epsilon\geq \frac{\log K_d}{\log N},\quad
\bar{\beta}+\epsilon\geq \frac{\log K}{\log N}.
\end{align}
The collection of all the achievable $(\bar{\alpha},\bar{\beta})$ pairs is the achievable region $\mathcal{R}$ of the $(4,3,3)$ exact-repair regenerating codes.
%Similarly, a normalized storage-bandwidth-tuple $(\bar{\alpha},\{\bar{\beta}_{i,{A},j}\})$ is said to be achievable for $(n,k,d)$ regenerating with unbalanced repair if for any $\epsilon>0$ there exists an $(n,k,d,N,K,\{K_{i,{A},j}\})$ code such that
%\begin{align}
%\bar{\alpha}+\epsilon\geq \frac{\log K}{\log N}
%\end{align}
%and
%\begin{align}
%\bar{\beta}_{i,{A},j}+\epsilon\geq \frac{\log K_{i,{A},j}}{\log N},\quad i\in{I}_n,\quad{A}\subseteq {I}_n\setminus\{i\}\quad \mbox{and}\quad |{A}|=d,\quad \, j\in {A}.
%\end{align}
\end{definition}

The reconstruction condition (\ref{eqn:reconstructionzeroerror}) requires that there is no decoding error, {\em i.e.}, the zero-error requirement is adopted. An alternative definition is to require instead the probability of decoding error to be asymptotically zero as $N\rightarrow \infty$. It will become clear from the rate region characterization given in the sequel that this does not cause any essential difference, and thus we do not give this alternative definition to avoid repetition.

\subsection{Some Further Notation}
In order to derive the outer bound, it is convenient to write the reconstruction and regenerating conditions in the form of entropy constraints. For this purpose, some further notation is introduced here, which is largely borrowed from \cite{RashmiShah:12:1}.

Let us denote the message random variable as $M$, which is uniformly distributed in the set $I_N$. Define
\begin{align}
W_i=f^E_i(M),\quad S_{i,j}=F^E_{i,j}\left(f^E_i(M)\right).
\label{WSdefinition}
\end{align}
Thus we have the following random variables in the set $\mathcal{W}\cup\mathcal{S}$
\begin{align}
\mathcal{W}=&\{W_1,W_2,W_3,W_4\},\label{eqn:Ws}\\
\mathcal{S}=&\{S_{1,2},S_{1,3},S_{1,4},S_{2,1},S_{2,3},S_{2,4},\nonumber\\
&\quad S_{3,1},S_{3,2},S_{3,4},S_{4,1},S_{4,2},S_{4,3}\}.\label{eqn:Ss}
\end{align}
The reconstruction requirement thus implies that
\begin{align}
H(\mathcal{W}\cup\mathcal{S}|\mathcal{A})=0,\quad \mbox{any } \mathcal{A}\subseteq \mathcal{W}: |\mathcal{A}|=3.
\label{eqn:reconstruction}
\end{align}
The regenerating requirement implies that
\begin{align}
H(S_{i,j}|W_i)=0,\quad j\in I_4,\quad i\in I_4\setminus \{j\},
\label{eqn:regeneratingencoding}
\end{align}
and 
\begin{align}
H(W_j|\{S_{i,j}\in \mathcal{S}: i\in I_n\setminus\{j\}\})=0,\quad \mbox{any } j\in I_4. 
\label{eqn:regenerating}
\end{align}
Because the message $M$ has a uniform distribution, we also have that
\begin{align}
H(\mathcal{W}\cup\mathcal{S})=H(M)=\log N\triangleq B,
\label{eqn:totalinfo}
\end{align}
which is strictly larger than zero. Note that together with (\ref{eqn:reconstruction}), this implies that
\begin{align}
H(\mathcal{A})=B,\quad \mbox{any } \mathcal{A} \mbox{ such that } |\mathcal{A}\cap\mathcal{W}|\geq 3.
\label{eqn:totalinfo2}
\end{align}
The symmetric storage requirement can be written as
\begin{align}
H(W_i)\leq \log K_d \triangleq \alpha,\quad W_i\in \mathcal{W},
\label{eqn:alpha}
\end{align}
and the regenerating bandwidth constraint can be written as
\begin{align}
H(S_{i,j})\leq \log K \triangleq \beta,\quad S_{i,j}\in \mathcal{S}.
\label{eqn:beta}
\end{align}
The above constraints (\ref{eqn:reconstruction})-(\ref{eqn:beta}) are the constraints that need to be satisfied by any exact-repair regenerating codes. These constraints will be used later in the converse proof.

\begin{figure}[tcb]
  \centering
  \includegraphics[width=.95\linewidth]{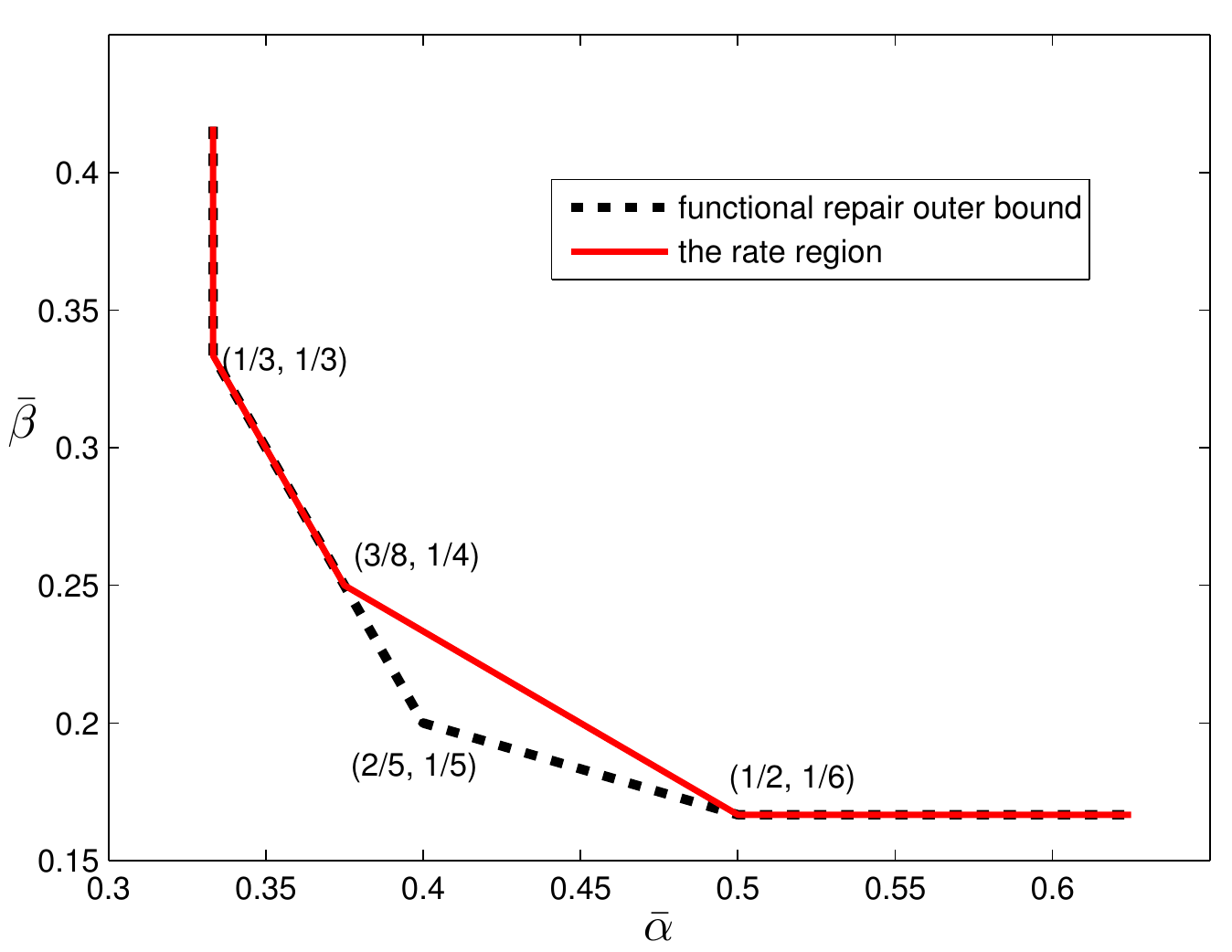}
  \caption{The functional-repair outer bound and the rate-region $\mathcal{R}$.\label{fig:bound433}}
\end{figure}

\subsection{Review of Functional-Repair Outer Bound}
\label{sec:review}
The optimal tradeoff for functional-repair regenerating codes was given by Dimakis {\em et al.} \cite{Dimakis:10}, which provides an outer bound for the exact-repair case. The bound has the following form in our notation for the $(4,3,3)$ case (see Fig. \ref{fig:bound433})
\begin{align}
\sum_{i=0}^2\min(\bar{\alpha},(3-i)\bar{\beta})\geq 1.
\end{align}
It is not difficult to show that it can be rewritten as the following four simultaneous linear bounds
\begin{align}
3\bar{\alpha}\geq 1,\quad 2\bar{\alpha}+\bar{\beta}\geq 1,\quad
\bar{\alpha}+3\bar{\beta}\geq 1,\quad 6\bar{\beta}\geq 1.\label{eqn:frouterbound}
\end{align}
The MSR point for this case is $(\bar{\alpha},\bar\beta)=(\frac{1}{3},\frac{1}{3})$, and the MBR point is  $(\bar{\alpha},\bar\beta)=(\frac{1}{2},\frac{1}{6})$.

\section{The Rate Region of $(4,3,3)$ Regenerating Codes}
\label{sec:main}

The following theorem provides a complete characterization of the rate region of the $(4,3,3)$ exact-repair regenerating codes.
\begin{theorem}
\label{theorem:main}
The rate region $\mathcal{R}$ of the $(n,k,d)=(4,3,3)$ exact-repair regenerating codes is given by the collection of $(\alpha,\beta)$ pairs that satisfy the following constraints
\begin{align*}
3\bar{\alpha}\geq 1,\quad 2\bar{\alpha}+\bar{\beta}\geq 1,\quad 4\bar{\alpha}+6\bar{\beta}\geq 3,\quad 6\bar{\beta}\geq 1.
\end{align*}
\end{theorem}

This rate region is also depicted in Fig. \ref{fig:bound433}, together with the functional-repair outer bound. It is clear that there is a gap between them, and thus the functional-repair outer bound cannot be asymptotically achievable under the exact-repair requirement. Note that the only difference between the region given in Theorem \ref{theorem:main} and that in (\ref{eqn:frouterbound}) is the third bounding plane. 

%The rest of the paper is devoted to proving this theorem.

\subsection{The Achievability Proof}
\label{sec:achievability}

The rate region $\mathcal{R}$ has three corner points, and thus we only need to show that these three points are all achievable. The MSR point $(\bar{\alpha},\bar\beta)=(\frac{1}{3},\frac{1}{3})$ is simply achieved by any $(4,3)$  MDS code, such as the binary systematic code with a single parity check bit. The MBR point $(\bar{\alpha},\bar\beta)=(\frac{1}{2},\frac{1}{6})$ is also easily obtained by using the repair-by-transfer code construction in \cite{RashmiShah:12:1}, which in this case reduces to a simple replication coding. It thus only remains to show that the point  $(\bar{\alpha},\bar\beta)=(\frac{3}{8},\frac{1}{4})$ is also achievable. 

Next we shall give a construction for a binary $(4,3,3)$ code with $\alpha=3$, $\beta=2$ and $B=8$, which indeed achieves this operating point. The code is illustrated in Table \ref{table:code}, where (and in the remainder of this section) the addition $+$ is in the binary field. Here $x_i,y_i,z_i,t_i$ are the systematic bits, $i=1,2$, and the remaining bits are the parity bits.

First note that the construction is circularly symmetric, and thus without loss of generality, we only need to consider the case when node 1 fails. If it can be shown that when node 2, 3, 4 each contribute two bits, node 1 can be reconstructed, which also implies that the complete data can be recovered using only node 2, 3 and 4, then the proof is complete. This can indeed be done using the combination shown in Table \ref{table:repair}.

Upon receiving these six bits  in Table \ref{table:repair}, the new node can form the following combinations
\begin{align*}
x_1+x_2&\quad+\quad y_1+y_2+z_1+\mathbf{z_2}+t_2\\
x_2&\quad+\quad y_1+y_2+z_1+z_2+t_1+\mathbf{t_1+t_2}\\
x_1&\quad+\quad \mathbf{y_1}+y_2+z_1+z_2+t_2,
\end{align*}
where the first combination is formed by using the second bit from node 2 and the first bit from node 3 (shown in bold), and the other combinations can be formed similarly. 
In the binary field, this is equivalent to having
\begin{align}
x_1+x_2&\quad+\quad y_1+y_2+z_1+z_2+t_2\label{eqn:first}\\
x_2&\quad+\quad y_1+y_2+z_1+z_2+t_2\label{eqn:second}\\
x_1&\quad+\quad y_1+y_2+z_1+z_2+t_2,\label{eqn:third}
\end{align} 
and it is seen that $x_1$ can be recovered by simply taking the difference between (\ref{eqn:first}) and (\ref{eqn:second}), and similarly $x_2$ can be recovered by taking the difference between (\ref{eqn:first}) and (\ref{eqn:third}). Note further that the third bit stored in node 1 is simply the summation of the first bits contributed from node 2, 3, and 4 in Table \ref{table:repair}. The proof is thus complete.

The hand-crafted code presented above is specific for the $(4,3,3)$ case. However, in a recent work, Sasidharan and Kumar \cite{SasidharanKumar:13} discovered a class of codes that is optimal for the $(n,n-1,n-1)$ case at operating points other than MSR or MBR, and specializing it to the $(4,3,3)$ case achieves the same performance as the code above; see also \cite{Tian:13-2} for a closely related code construction.

\begin{table}
\centering
\caption{A $(4,3,3)$ code for  $(\bar{\alpha},\bar\beta)=(\frac{3}{8},\frac{1}{4})$.}
\label{table:code}
\begin{tabular}{|c|c|c|c|}
\hline
&first bit&second bit&third bit\\\hline
node 1&$x_1$&$x_2$&$y_1+z_2+t_1+t_2$\\\hline
node 2&$y_1$&$y_2$&$z_1+t_2+x_1+x_2$\\\hline
node 3&$z_1$&$z_2$&$t_1+x_2+y_1+y_2$\\\hline
node 4&$t_1$&$t_2$&$x_1+y_2+z_1+z_2$\\\hline
\end{tabular}
\end{table}

\begin{table}
\centering
\caption{Repair contributions when node 1 fails.}
\label{table:repair}
\begin{tabular}{|c|c|c|}
\hline
&first bit&second bit\\\hline
node 2&$y_1$&$z_1+t_2+x_1+x_2+y_1+y_2$\\\hline
node 3&$z_2$&$t_1+x_2+y_1+y_2+z_1+z_2$\\\hline
node 4&$t_1+t_2$&$x_1+y_2+z_1+z_2+t_2$\\\hline
\end{tabular}
\end{table}

\subsection{The Converse Proof}
\label{sec:converse}

It is clear that we only need to prove the following bound
\begin{align}
4\bar{\alpha}+6\bar{\beta}\geq 3,
\end{align}
because the other bounds in the main theorem can be obtained from the outer bound (\ref{eqn:frouterbound}). We first give an instrumental result regarding the symmetry of the optimal solution.

\begin{definition}
A permutation $\pi$ on the set $I_4$ is a one-to-one mapping $\pi:{I}_4\rightarrow {I}_4$. The collections of all permutations is denoted as $\Pi$.
\label{def:permutation}
\end{definition}

Any given permutation $\pi$ correspondingly maps a random variable $W_i$ to $W_{\pi(i)}$.  Any subset of $\mathcal{W}$, e.g., $\mathcal{A}\subseteq \mathcal{W}$, is thus mapped to another set of random variables, denoted as $\pi(\mathcal{A})$. For example, the permutation $\pi(1)=2$, $\pi(2)=3$, $\pi(3)=1$ and $\pi(4)=4$ will map the set of random variables $\mathcal{A}=\{W_1,W_4\}$ to $\pi(\mathcal{A})=\{W_2,W_4\}$. Similarly a random variable $S_{i,j}$ will be mapped to $S_{\pi(i),\pi(j)}$, and for any subset of $\mathcal{S}$, we use a similar notation as for the case of $\mathcal{W}$.

\begin{definition}
\label{def:symmetricentropy}
An $(N,K_d,K)$ exact-repair regenerating code is said to induce a symmetric entropic vector if for any sets $\mathcal{A}\subseteq \mathcal{S}$ and $\mathcal{B}\subseteq\mathcal{W}$ and any permutation $\pi\in\Pi$, 
\begin{align}
H(\mathcal{A},\mathcal{B})=H(\pi(\mathcal{A}),\pi(\mathcal{B})).
\end{align}
\end{definition}

\begin{definition}
A normalized bandwidth-storage pair $(\bar{\alpha},\bar{\beta})$ is said to be symmetrically $(4,3,3)$ exact-repair achievable if for any $\epsilon>0$ there exists an $(N,K_d,K)$ exact-repair regenerating code which induces a symmetric entropic vector such that
\begin{align}
\bar{\alpha}+\epsilon\geq \frac{\log K_d}{\log N},\quad
\bar{\beta}+\epsilon\geq \frac{\log K}{\log N}.
\end{align}
The collection of all such $(\bar{\alpha},\bar{\beta})$ pairs is the entropy-symmetrical achievable region $\mathcal{R}^*$ of the $(4,3,3)$ exact-repair regenerating codes.
\end{definition}

With the above definition, it is not difficult to see that the following proposition is true.
\begin{prop}
\label{prop:symmetry}
For $(n,k,d)=(4,3,3)$ exact-repair regenerating codes $\mathcal{R}=\mathcal{R}^*$.
\end{prop}

Clearly the inclusion $\mathcal{R}^*\subseteq\mathcal{R}$ is true. For the other direction, we can invoke a time-sharing (or more precisely here, space-sharing) argument among all possible permutations; the proof is given in the appendix for completeness. We are now ready to prove the converse of Theorem \ref{theorem:main}.

%\subsection{Converse Proof of Theorem 1}
%\allowdisplaybreaks

\begin{IEEEproof}[Converse Proof of Theorem \ref{theorem:main}]
Because of the equivalence in Proposition \ref{prop:symmetry}, without loss of generality we only need to prove that the given outer bound holds for the region $\mathcal{R}^*$, {\em i.e.}, consider only codes that induce symmetrical entropy vectors. 

We first write\footnote{The proof presented here is different from the one given in the preliminary conference version for the same result \cite{Tian:ISIT13}. The proof here is more concise because  further reduction has been applied in the secondary LP problem discussed in the next section.}
\begin{align}
4\alpha + 6\beta \geq& 4H(W_1)+6H(S_{2,4})\nonumber\\
=& 4H(W_1)+3H(S_{2,4})+3H(S_{3,4})\nonumber\\
\geq& H(W_1)+3H(W_1S_{2,4}S_{3,4})
%\geq& 4H(S_{3,1}S_{2,1}W_4)+4H(S_{3,2}W_4)\nonumber\\
%&-2I(S_{2,1};W_3|W_4)-2I(W_3;W_4|S_{3,2}S_{3,4}S_{2,4})\nonumber\\
%=&4H(S_{3,1}S_{2,1}W_4)+2H(W_4)+2H(W_3W_4S_{2,1})\nonumber\\
%&-2H(W_3W_4)+2H(S_{3,2}S_{3,4}S_{2,4})\nonumber\\
%&+2H(W_3W_4S_{2,4}S_{3,2}S_{3,4})-2H(W_4S_{3,2}S_{3,4}S_{2,4})\nonumber\\
%=&4H(S_{3,1}S_{2,1}W_4)+2H(W_4)+2B\nonumber\\
%&-2H(W_3W_4)+2H(S_{3,2}S_{3,4}S_{2,4})\nonumber\\
%&+2H(W_3W_4S_{2,4})-2H(W_4S_{3,2}S_{3,4}S_{2,4})
\end{align}
where the first inequality is by (\ref{eqn:alpha}) and (\ref{eqn:beta}), the equality is by the symmetry
of the solution
\begin{align}
H(S_{2,4})=H(S_{3,4}),
\end{align}
and the second inequality is because summation of individual entropy is greater than or equal to the joint entropy. 

%Notice that
%\begin{align}
%H(W_1S_{2,4}S_{3,4})=&H(W_1S_{1,4}S_{2,4}S_{3,4})
%&=H(W_1W_4S_{2,4}S_{3,4})\nonumber\\
%\geq&H(W_1W_4S_{2,4})
%\end{align}
%where the first equality is by applying (\ref{eqn:regeneratingencoding}) and the second equality by (\ref{eqn:regenerating}), and we can write

For notational simplicity,  from here on we shall write (s), (\ref{eqn:reconstruction}), (\ref{eqn:regeneratingencoding}), (\ref{eqn:regenerating}) and (\ref{eqn:totalinfo2}) on top of the equalities in the derivation to signal the reasons for the equalities, {\em i.e.,} by the symmetry of the entropy vectors, or by equations (\ref{eqn:reconstruction}), (\ref{eqn:regeneratingencoding}), (\ref{eqn:regenerating}) and (\ref{eqn:totalinfo2}), respectively. 
We next write the following chain of inequalities
\begin{align}
&2H(W_1S_{2,4}S_{3,4})\nonumber\\
&\stackrel{(\ref{eqn:regeneratingencoding})}{=}2H(W_1S_{1,4}S_{2,4}S_{3,4})\nonumber\\
&\stackrel{(\ref{eqn:regenerating})}{=}2H(W_1W_4S_{1,4}S_{2,4}S_{3,4})\nonumber\\
&\stackrel{(\ref{eqn:regeneratingencoding})}{=}2H(W_1W_4S_{2,4}S_{3,4})\nonumber\\
&\geq H(W_1W_4S_{2,4})+H(W_1W_4S_{2,4}S_{3,4})\nonumber\\
&\stackrel{(s)}{=}H(W_2W_4S_{1,2})+H(W_1W_4S_{2,4}S_{3,4})\nonumber\\
&\stackrel{(\ref{eqn:regeneratingencoding})}{=}H(W_2W_4S_{1,2}S_{2,4})+H(W_1W_4S_{1,2}S_{2,4}S_{3,4})\nonumber\\
&=H(W_2|W_4S_{1,2}S_{2,4})+H(W_1S_{3,4}|W_4S_{1,2}S_{2,4})\nonumber\\
&\qquad\qquad+2H(W_4S_{1,2}S_{2,4})\nonumber\\
&\geq H(W_1W_2S_{3,4}|W_4S_{1,2}S_{2,4})+2H(W_4S_{1,2}S_{2,4})\nonumber\\
&=H(W_1W_2W_4S_{1,2}S_{2,4}S_{3,4})+H(W_4S_{1,2}S_{2,4})\nonumber\\
&\stackrel{(\ref{eqn:totalinfo2})}{=}B+H(W_4S_{1,2}S_{2,4}).
\end{align}
It follows that
\begin{align}
&4\alpha + 6\beta \nonumber\\
&\quad\geq B+H(W_1)+H(W_1S_{2,4}S_{3,4})+H(W_4S_{1,2}S_{2,4}).
\end{align}

However, notice that
\begin{align}
&H(W_1S_{2,4}S_{3,4})+H(W_4S_{1,2}S_{2,4})\nonumber\\
&\stackrel{(\ref{eqn:regeneratingencoding})}{=}H(W_1S_{1,4}S_{2,4}S_{3,4})+H(W_4S_{1,2}S_{2,4})\nonumber\\
&\stackrel{(\ref{eqn:regenerating})}{=}H(W_1W_4S_{1,4}S_{2,4}S_{3,4})+H(W_4S_{1,2}S_{2,4})\nonumber\\
&\stackrel{(s)}{=}H(W_1W_4S_{1,4}S_{2,4}S_{3,4})+H(W_4S_{3,2}S_{2,4})\nonumber\\
&=H(W_1S_{1,4}S_{3,4}|W_4S_{2,4})+H(S_{3,2}|W_4S_{2,4})\nonumber\\
&\qquad\qquad+2H(W_4S_{2,4})\nonumber\\
&\geq H(W_1S_{1,4}S_{3,2}S_{3,4}|W_4S_{2,4})+2H(W_4S_{2,4})\nonumber\\
&=H(W_1W_4S_{1,4}S_{2,4}S_{3,2}S_{3,4})+H(W_4S_{2,4})\nonumber\\
&\stackrel{(\ref{eqn:regeneratingencoding})}{=}H(W_1W_4S_{1,2}S_{3,2}S_{4,2}S_{1,4}S_{2,4}S_{3,4})+H(W_4S_{2,4})\nonumber\\
&\stackrel{(\ref{eqn:regenerating})}{=}H(W_1W_2W_4S_{1,2}S_{3,2}S_{4,2}S_{1,4}S_{2,4}S_{3,4})+H(W_4S_{2,4})\nonumber\\
&\stackrel{(\ref{eqn:totalinfo2})}{=}B+H(W_4S_{2,4}).
\end{align}
This implies that
\begin{align}
&4\alpha + 6\beta \nonumber\\
&\geq 2B+H(W_1)+H(W_4S_{2,4})\nonumber\\
&\stackrel{(s)}{=} 2B+H(W_2)+H(W_4S_{2,4})\nonumber\\
&=2B+H(W_2)+H(S_{3,1})+H(W_4S_{2,4})-H(S_{3,1})\nonumber\\
&\geq 2B+H(W_2S_{3,1})+H(W_4S_{2,4})-H(S_{3,1})\nonumber\\
&\stackrel{(\ref{eqn:regeneratingencoding})}{=}2B+H(W_2S_{2,4}S_{3,1})+H(W_4S_{2,4})-H(S_{3,1})\nonumber\\
&\stackrel{(s)}{=}2B+H(W_2S_{2,4}S_{3,1})+H(W_4S_{2,4})-H(S_{2,4})\nonumber\\
&=2B+H(W_2S_{3,1}|S_{2,4})+H(W_4|S_{2,4})+H(S_{2,4})\nonumber\\
&\geq 2B+H(W_2W_4S_{3,1}|S_{2,4})+H(S_{2,4})\nonumber\\
&=2B+H(W_2W_4S_{3,1}S_{2,4})\nonumber\\
&\stackrel{(\ref{eqn:regeneratingencoding})}{=}2B+H(W_2W_4S_{2,1}S_{3,1}S_{4,1}S_{2,4})\nonumber\\
&\stackrel{(\ref{eqn:regenerating})}{=}2B+H(W_1W_2W_4S_{2,1}S_{3,1}S_{4,1}S_{2,4})\nonumber\\
&\stackrel{(\ref{eqn:totalinfo2})}{=}3B,
\label{eqn:finalstep}
\end{align}
and the proof is thus complete.
\end{IEEEproof}

It can be seen that the rate region given here remains the same if the codes are required only to have asymptotic zero error probability as $B$ approaches infinity, instead of the more stringent zero-error requirement. We only need to replace the steps where (\ref{eqn:totalinfo2}) was applied in the above proof to a slightly different version based on Fano\rq{}s inequality \cite{CoverThomas}, and the details are thus omitted. 

\section{The Computer-Aided Proof Approach}
\label{sec:CAP}

It should be clear at this point that the converse proof given above is difficult to find manually, in which several rather unconventional steps, such as the adding and subtracting of the same term in the third step of (\ref{eqn:finalstep}), are used. These steps arise from the complex symmetry and dependence structure in the random variables. In fact, the author\rq{}s own extensive  attempt to find such a converse proof manually was utterly unsuccessful, which motivated the investigation of the computer-aided proof (CAP) approach, via which the proof was obtained. In this section, we provide details on our approach used to obtain this converse, which may prove useful for future research.

\subsection{The Basic Primal Linear Program}

Yeung provided a linear programing framework to prove Shannon-type information inequalities \cite{Yeung:book}. The basic framework can be roughly described as follows in the context of the problem being considered. 

There are a total number of $2^{16}-1=65535$ joint entropy terms, each of which corresponds to a non-empty subset of the set of $16$ random variables in the problem being considered [see (\ref{eqn:Ws}) and (\ref{eqn:Ss})]. Note that every information measure (entropy, conditional entropy, mutual information and conditional mutual information) can be represented as a linear combination of these joint entropies. These joint entropy terms can be viewed as the variables in an optimization problem, and they must satisfy the constraints imposed by the problem, as well as the more general non-problem-specific constraints. 

One particular set of general constraints on joint entropies are those imposed by the non-negativity of Shannon\rq{}s information measures, which are collectively called Shannon-type inequalities \cite{Yeung:book}. It was shown that without loss of generality, this set of constraints can be represented by the following minimal set of constraints for a collection of random variables $\mathcal{X}=\{X_1,X_2,\ldots,X_n\}$:
\begin{align}
&H(X_i|\{X_k, k\neq i\})\geq 0,\quad i\in I_{n}\label{eqn:Shannontype1}\\
&I(X_i;X_j|\{X_k, k\in K\})\geq 0, \, \mbox{where }  k\in I_n-\{i,j\},\, i\neq j.\label{eqn:Shannontype2}
\end{align}
Note that in our problem $\mathcal{X}=\mathcal{S}\cup \mathcal{W}$. It can be seen that there are a total of $16+{16 \choose 2}2^{14}=1966096$ constraints in the problem being considered. 

Since all the constraints are linear, without loss of generality, we can set $B=1$. When sweeping through $\alpha$, for each instance we shall fix $\alpha=\alpha_0\in [\frac{3}{8},\frac{1}{2}]$, and seek to find the minimum value of $\beta$ without violating these constraints, {\em i.e.}, a lower bound for $\beta$ when $\alpha=\alpha_0$ and $B=1$. In addition to the $65535$ variables representing the joint entropy terms, an auxiliary variable $\beta$ is also introduced in the linear program, which will be the objective function subject to minimization. More precisely, we shall consider the following basic optimization problem:
\begin{align}
\mbox{minimize: }&\beta\nonumber\\
\mbox{subject to: }
& B=1\nonumber\\
& \alpha=\alpha_0\nonumber\\
&\mbox{constrains (\ref{eqn:Shannontype1})-(\ref{eqn:Shannontype2})}\nonumber\\
&\mbox{constraints (\ref{eqn:reconstruction})-(\ref{eqn:regenerating}), (\ref{eqn:totalinfo})-(\ref{eqn:beta})}.
\end{align}
Note that (\ref{eqn:totalinfo2}) are redundant constraints given (\ref{eqn:reconstruction}) and (\ref{eqn:totalinfo}) and the Shannon-type inequalities, and thus it is not included in the minimization problem. In this optimization problem, there are a total $1966096+4+12=1966112$ inequality constraints from the Shannon-type inequalities and the constraints (\ref{eqn:alpha})-(\ref{eqn:beta}), as well as $4+12+4+1=21$ equality constraints from (\ref{eqn:reconstruction})-(\ref{eqn:regenerating}) and (\ref{eqn:totalinfo}), after substituting the values of $B$ and $\alpha$ into the constraints. It is worth noting that the solution for the above optimization problem is potentially only an outer bound for the rate region, since it does not take into account of non-Shannon type inequalities; nevertheless, it turns out for this problem Shannon-type inequalities are in fact sufficient.  

This optimization problem in its basic form is too large for the existing software packages using Yeung\rq{}s LP framework, {\em i.e.}, Information Theoretic Inequality Prover (ITIP or XITIP) \cite{ITIP}\cite{XITIP}. In fact, LP problems at this scale in general, with the large total number of constraints in this problem in particular, are on the border of the problems that modern commercial optimization solvers are able to handle. Depending on the software library being used, it may take a few hundred hours without converging. For example, when the popular Cplex optimization library \cite{Cplex} is used, running the solver for 48 hours (2-thread mode on a multi-core Linux server of CPU frequency at 2.66GHz) does not yield a solution for the problem in the basic form, and  for any practical purpose it is safe to deem the problem in this form too complex for the solver. 

\subsection{Dimension Reduction in the Primal Linear Program}

To reduce the dimension of the LP problem, we take advantage of the symmetry and other problem specific structure, as detailed in the sequel. 

Firstly, by the existence of the optimal symmetric solution ({\em i.e.}, with a symmetric entropic vector), the number of variables in the LP problem can be reduced. For example, the variable representing the entropy term $H(W_1,S_{2,3})$ has the same value as any variables representing the entropy terms of the form $H(W_i,S_{j,k})$, where $i, j, k$ are distinct elements of $I_4$. Thus these variables in the LP problem can be eliminated except one arbitrary member of them. In the above example, if a variable representing a joint entropy of the form $H(W_i,S_{j,k})$ appears in any equality or inequality constraints, it can be replaced by the variable representing $H(W_1,S_{2,3})$.

Secondly, recall the equality (\ref{eqn:totalinfo2}), which is implied by (\ref{eqn:reconstruction}) and (\ref{eqn:totalinfo}) through the application of the Shannon-type inequalities in (\ref{eqn:Shannontype2}) and (\ref{eqn:Shannontype1}). This implies that the variables in the LP problem representing any joint entropy $H(\mathcal{A})$ such that
$|\mathcal{A}\cap\mathcal{W}|\geq 3$ is of value $B$. More generally, consider the following set growth procedure for a set of random variables $\mathcal{A}\subseteq \mathcal{S}\cup\mathcal{W}$:
\begin{enumerate}
\item Initialize $\mbox{gr}(\mathcal{A})=\mathcal{A}$;
\item For $i=1,2,3,4$: if $W_i\in \mbox{gr}(\mathcal{A})$, let $\mbox{gr}(\mathcal{A})=\mbox{gr}(\mathcal{A})\cup \{S_{i,j},S_{i,j},S_{i,t}\}$ where $j,k,t$ are distinct elements of $I_4$ and not equal to $i$;
\item For $i=1,2,3,4$: if $\{S_{j,i},S_{k,i},S_{t,i}\}\subseteq \mbox{gr}(\mathcal{A})$, let $\mbox{gr}(\mathcal{A})=\mbox{gr}(\mathcal{A})\cup \{W_i\}$, where $j,k,t$ are distinct elements of $I_4$ and not equal to $i$;
\item If the set $\mbox{gr}(\mathcal{A})$ did not grow in the previous two steps, exit; otherwise, return to step (2). 
\end{enumerate}
It is clear that if the resultant set $\mbox{gr}(\mathcal{A})$ satisfies $|\mbox{gr}(\mathcal{A})\cap\mathcal{W}|\geq 3$, then $H(\mathcal{A})$ is also of value $B$. Thus the corresponding variables in the LP problem can be eliminated. 

%Moreover, consider the following reduction procedure for a set of random variables $\mathcal{A}\subseteq \mathcal{S}\cup\mathcal{W}$:
%\begin{enumerate}
%\item Initialize $\mbox{re}(\mathcal{A})=\mathcal{A}$;
%\item For $i=1,2,3,4$: if $W_i\in \mbox{re}(\mathcal{A})$, let $\mbox{re}(\mathcal{A})=\mbox{re}(\mathcal{A})\setminus \{S_{i,j},S_{i,j},S_{i,t}\}$ where $j,k,t$ are distinct elements of $I_4$ and not equal to $i$;
%\item For $i=1,2,3,4$: if $\{S_{j,i},S_{k,i},S_{t,i}\}\subseteq\mbox{re}(\mathcal{A})$, let $\mbox{re}(\mathcal{A})=\mbox{re}(\mathcal{A})\setminus \{W_i\}$, where $j,k,t$ are distinct elements of $I_4$ and not equal to $i$. 
%\end{enumerate}
Moreover, it is clear that the if any two subsets of $\mathcal{S}\cup\mathcal{W}$, denoted as $\mathcal{A}$ and $\mathcal{B}$, satisfy $\mbox{gr}(\mathcal{A})=\mbox{gr}(\mathcal{B})$, then $H(\mathcal{A})=H(\mathcal{B})$. In other words, these subsets form an equivalent class, and the variables representing them in the LP problem can be eliminated except a single arbitrary member of them. Furthermore, after utilizing the equivalent class relation, the equality constraints (\ref{eqn:reconstruction})-(\ref{eqn:regenerating}) and (\ref{eqn:totalinfo}) can now be completely eliminated.

Without loss of optimality, the inequality (\ref{eqn:alpha}) can be taken to be equality, and thus the variables corresponding to $H(W_i)$ can be eliminated; similarly the inequality (\ref{eqn:beta}) can also be taken to be equality, and thus the variable corresponds to $H(S_{i,j})$ can also be eliminated and replaced by $\beta$. 

Lastly, with the above reductions of variables in the LP problem, many inequality constraints become degenerate ({\em i.e.},  in the form of $0\geq 0$) or repetition of others, and they can be removed from the set of constraints. 

After the above steps, there remain in the LP problem only a total of $176$ variables, and a total of $6152$ inequality constraints, which ({\em i.e.}, each instance with a fixed $\alpha_0$ value) can be solved in less than $0.1$ second using the Cplex solver under the same setup as previously mentioned. 
\subsection{The Secondary LP}

The reduced primal LP problem allows us to trace out a lower bound for the exact-repair regenerating code rate region {\em numerically}. There are two concerns for this numeric approach: it is not clear how many values of $\alpha$ we should choose to accurately trace out the rate region, and the solution is numerical which implies that the bound such obtained is only accurate within numerical precision.

There are various methods to address these concerns, however, we wish to find an explicit information theoretic proof (algebraic proof) for its obvious conceptual advantage. Let us consider a hypothetical bound that
\begin{align}
\gamma_\alpha \alpha+\gamma_\beta \beta \geq \gamma_B B,
\end{align}
and moreover, the coefficients $(\gamma_\alpha, \gamma_\beta,\gamma_B)$ are chosen such that there exists $(\alpha,\beta,B)$ triples that satisfy it with equality under the constraints in the primal problem given in the previous sub-section. This assumption is equivalent to saying that this bound is the tightest outer bound that can be obtained under these constraints.

Recall that the reduced primal LP problem has $176$ variables, and $6152$ constraints. Let us consider an expanded version of the optimization problem where:
\begin{itemize}
\item $\alpha$ and $B$ are also variables;
\item The objective function subject to minimization is $\gamma_\alpha \alpha+\gamma_\beta \beta - \gamma_B B$;
\item The constraint $B=1$ is rewritten as follows (as the first two constraints when the optimization problem is written in a matrix form)
\begin{align}
B\leq 1, \quad -B\leq -1;
 \label{eqn:Bs}
\end{align}
\item The constraint $\alpha=\alpha_0$ is removed.
\end{itemize}

%If the numerical bound obtained using the procedure above is indeed the tightest that we can obtain under the given constraints, then the minimization problem specified above has a solution of zero. 
Under the assumption aforementioned, this minimization problem specified above has a solution of zero. 
The reason to let $B=1$ is to avoid the optimal but less meaningful all-zero solution, and we wrote it as two inequalities such that the minimization problem has the following standard LP form, where $(B,\alpha,\beta)$ correspond to variables $(x_1,x_2,x_3)$ and the $175$ entropies terms (in the reduced primal LP problem) correspond to the variables $x_{4},x_5,\ldots,x_{178}$:
\begin{align}
\mbox{minimize: }& c^t\vec{x}\nonumber\\
\mbox{subject to: }& A\vec{x}\leq b, \label{eqn:secondprimal}
\end{align}
where $b$ and $c$ are both column vectors given as 
\begin{align}
b&=[1, -1, 0, 0, \ldots, 0]^t\nonumber\\
c&=[-\gamma_B, \gamma_\alpha, \gamma_\beta, 0, 0,\ldots,0]^t.
\end{align}

The standard dual problem is thus given by (see \cite{Boydbook})
\begin{align}
\mbox{maximize: }& -b^t\vec{\lambda}\nonumber\\
\mbox{subject to: }& A^t\vec{\lambda}+c=0\label{eqn:dual}\\
&\vec{\lambda}\geq 0.
\end{align}

Note here $\vec{\lambda}$ is a vector of length $6152$. Clearly the primal LP problem has a trivial solution that all joint entropy terms are equal to $B_0$, and thus it is feasible. This fact further implies that the stronger duality holds in this case, and there is no duality gap between the primal and dual problem \cite{Boydbook}. 
The solution for the dual problem is not unique, but because the primal LP problem has solution zero, any such solution must satisfy
\begin{align}
-b^t\vec{\lambda} = 0,
\end{align}
which implies that $\lambda_1=\lambda_2$. From (\ref{eqn:dual}), it is seen then that $(\lambda_{3},\lambda_4,\ldots,\lambda_{6152})$ in any optimal solution for the dual problem leads to an explicit proof of the inequality $\gamma_\alpha \alpha+\gamma_\beta \beta \geq \gamma_B B$. This is because each column of the matrix $A^t$ is a known information inequality, {\em i.e.}, either a Shannon-type inequality, or an inequality implied by the combination of Shannon-type inequalities and certain problem-specific constraints; therefore, any non-negative linear combination is a valid information inequality, where the linear coefficients gives an explicit proof of the resultant inequality. 

Let $A_o$ be the submatrix of $A$ which does not include the first two rows, {\em i.e.}, removing the constraints corresponding to $B=1$. The above discussion essentially states that there exists a vector $\vec{\lambda}_o$ such that $A^t_o\vec{\lambda}_o+c=0$ for any bound that satisfy the tightness assumption, and it gives an explicit information theoretic proof; in other words, using $\vec{\lambda}_o$ as the weights in the linear combination of the known information inequalities, we obtain the desired inequality. 

Clearly from the procedure discussed in the previous section, we wish to prove $4\alpha+6\beta\geq 3B$, and for this purpose, we only need to solve the equality $A^t_o\vec{\lambda}_o+c=0$, where 
\begin{align}
c&=[-3, 4, 6, 0, 0,\ldots,0]^t,
\end{align}
and the solution must exists if $4\alpha+6\beta\geq 3B$ is the tightest bound under the primal LP constraints.

\begin{table}[tcb]
\begin{center}
\caption{The non-zero terms in the secondary LP solution.}
\label{tab:correspondence}
\begin{tabular}{|c|c|}
\hline
$y_1$&$H(W_iS_{j,i})$\\
$y_2$&$H(W_iS_{j,k})$\\
$y_3$&$H(W_iS_{j,i}S_{k,i})$\\
$y_4$&$H(W_iS_{j,i}S_{k,j})$\\
$y_5$&$H(W_iS_{j,k}S_{t,k})$\\
$y_6$&$H(W_iS_{j,i}S_{j,k}S_{k,i})$\\
$y_7$&$H(W_iW_j)$\\
$y_8$&$H(W_iW_jS_{k,i})$\\
\hline
\end{tabular}
\end{center}
\end{table}

The solution for this set of equations is not unique, and here we wish to find one that has the fewest non-zero elements ({\em i.e.}, sparsest solution of $\vec{\lambda}_o$), which roughly translates to a converse proof with the fewest derivation steps. This $\ell_0$ optimization problem is however NP-hard, but it is well known that the $\ell_1$ norm can be used to approximate the  $\ell_0$ norm, and thus we can solve the alternative optimization problem:
\begin{align}
\mbox{maximize: }& \sum \lambda_i\nonumber\\
\mbox{subject to: }& A_o^t\vec{\lambda}_o+c=0\nonumber\\
&\vec{\lambda}\geq 0\nonumber.
\end{align}
This problem is an LP problem, and by solving this secondary LP problem, we directly obtain an explicit information theoretic solution. It should be emphasized that the  approximation is in the sense that the solution thus obtained is not the sparsest solution that we desired, but not that the proof is of an approximate nature: any solution of the equalities $A^t_o\vec{\lambda}_o+c=0$ is a valid explicit information theoretic converse proof, even if it is not the sparsest.

\begin{table}[tc]
\setlength{\tabcolsep}{4pt}
\begin{center}
\caption{The solution for the dual problem}
\label{table:cancellation}
\begin{tabular}{|ccccccccccc|}
\hline
$B$&$\alpha$&$\beta$&$y_1$&$y_2$&$y_3$&$y_4$&$y_5$&$y_6$&$y_7$&$y_8$\\\hline\hline
      &$7$       &$7$     &          & $-7$  &         &         &         &         &         &              \\
      &$-3$      &           &           &$ 6$  &         &         &$-3$  &         &        &       \\
      &$1$       & $-1$   & $1$     &         &         &         &         &         & $-1$&       \\
 $-1$& $-1$    &           &           &  $1$ &         &         &         &         &  $1$&       \\
$-1$ &            &           &           &         &         & $-1$ &  $ 1$&         &         & $1$      \\
 $-1$&            &           &           &         & $-1$ &         & $1$  & $1$   &         &       \\
      &            &           & $-1$   &         & $1$   &  $1$  &         & $-1$ &         &       \\
      &            &           &           &         &         &         &   $1$ &         &         & $-1$ \\\hline \hline
$-3$&    $4$  &  $6$   &           &         &         &         &         &          &        &        \\\hline     
\end{tabular}
\end{center}
\end{table}

\begin{table}[t]
\begin{center}
\caption{Rewriting the solution for the dual problem}
\label{table:basicinequalities}
\begin{tabular}{|c|c|}
\hline
Coefficients& Inequalities\\\hline
$7$&$I(S_{i,j};W_k)\geq0$\\
$3$&$I(S_{k,j};S_{t,j}|W_i)\geq0$\\
$1$&$I(W_i;W_j|S_{i,j})\geq0$\\
$1$&$I(W_i;S_{t,k}|W_j)\geq0$\\
$1$&$I(W_i;W_jS_{k,t}|S_{i,t}S_{j,i}W_t)\geq0$\\
$1$&$I(W_i;S_{k,t}|S_{k,j}S_{t,j}W_j)\geq0$\\
$1$&$I(S_{k,i};S_{k,j}|S_{j,i}W_i)\geq0$\\
$1$&$H(S_{t,i}|S_{k,i}W_iW_j)\geq0$\\
\hline
\end{tabular}
\end{center}
\end{table}

Only a small subset of the joint entropy terms are in the solution of this secondary LP problem, as listed in Table \ref{tab:correspondence} where we also given them labels $y_1,y_2,\ldots,y_8$ to facilitate subsequent discussion. Here the letters $i,j,k,t$ are used to denote four distinct indices in the set $I_4$, because by the symmetry, they may assume any order.
%\begin{table}
%\begin{center}
%\caption{Renaming the joint entropy terms in the secondary LP solution.}
%\label{tab:correspondence}
%\begin{tabular}{|c|c|}
%\hline
%$y_1$&$H(S_{i,j}S_{k,j})$\\
%$y_2$&$H(S_{i,j}S_{k,j}S_{i,k})$\\
%$y_3$&$H(S_{i,j}S_{k,j}S_{t,j}S_{i,k})$\\
%$y_4$&$H(S_{i,j}S_{k,j}S_{t,j}S_{i,k}S_{k,i})$\\
%$y_5$&$H(W_iS_{j,k})$\\
%$y_6$&$H(W_iS_{j,k}S_{t,k})$\\
%$y_7$&$H(W_iS_{j,i}S_{j,k}S_{t,k})$\\
%$y_8$&$H(W_iS_{j,i}S_{j,k}S_{k,i})$\\
%$y_9$&$H(W_iW_j)$\\
%$y_{10}$&$H(W_iW_jS_{k,i})$\\
%$y_{11}$&$H(W_iW_jS_{k,i}S_{t,i})$\\
%$y_{12}$&$H(W_iW_jS_{k,i}S_{t,j})$\\
%\hline
%\end{tabular}
%\end{center}
%\end{table}

We can now tabulate the solution of the secondary LP problem, as given in Table. \ref{table:cancellation}, one row corresponding to one row in $A_o$, {\em i.e.}, one basic information inequality as shown in Table \ref{table:basicinequalities}. The last line in Table. \ref{table:cancellation} is the row summation which is indeed $4\alpha+6\beta\geq 3B$.
Note that the last inequality of Table \ref{table:basicinequalities} is also a basic information inequality, but it is not in the form of (\ref{eqn:Shannontype2}) because some problem specific reduction discussed in the previous sub-section has been incorporated.  Though this is already a valid proof, we can manually combine several inequalities to simplify the proof, and the converse proof given in the previous section is the result after such further manual simplifications.

In \cite{Yeung:book}, Yeung showed that all unconstrained Shannon-type inequalities are linear combinations of elemental Shannon-type inequalities, {\em i.e.},  (\ref{eqn:Shannontype1}) and (\ref{eqn:Shannontype2}). The approach we have discussed above can be viewed as a generalization of this result under additional problem-specific constraints. However, the introduction of the $\ell_1$ norm objective function to approximately find the sparsest linear combination has not been used previously to investigate information inequalities, and thus it is a novel ingredient. Moreover, the proof given in \cite{Yeung:book} relies on the fact that all joint entropies can be represented by a linear combinations of the elementary forms of Shannon\rq{}s information measures, which are the left hand sides of (\ref{eqn:Shannontype1}) and (\ref{eqn:Shannontype2}). Since our result is regarding the \textit{tightest} bounds that can be obtained using the LP approach, the proof directly follows from the strong duality without relying on the completeness of the elementary forms of Shannon\rq{}s information measures.

\section{Conclusion}
\label{sec:conclusion}

A complete characterization is provided for the rate region of the $(4,3,3)$ exact-repair regenerating codes, which shows that the cut-set outer bound \cite{Dimakis:10} is in general not (even asymptotically) tight for exact-repair. An explicit binary code construction is provided to show that the given rate region is achievable. One main novelty of the work is that a computer-aided proof approach is developed by extending Yeung\rq{}s linear programming framework, and an explicit information theoretic proof is directly obtained using this approach. We believe customizing the LP approach to other communication problems based on similar reduction techniques can be a rather fruitful path, which appears particularly suitable for research on storage systems, and thus have presented some related details in this work. 

Although sparsity is used approximately as an objective in the secondary LP problem, this sparsity is only with respect to the elementary Shannon-type inequalities (\ref{eqn:Shannontype1})-(\ref{eqn:Shannontype2}), and thus including more redundant basic Shannon-type inequalities may lead to even sparser solution. This is already evident in the algebraic proof given where manual simplification was taken and some basic inequalities not in (\ref{eqn:Shannontype1})-(\ref{eqn:Shannontype2}) were used. Including these basic inequalities in the secondary LP will clearly yield a more sparser solution. It should also be noted that sparsity only translates roughly to a small number of proof steps, but does not necessarily lead to a structured proof that can be extended to general parameter settings.

The result presented in this work revealed that the cut-set outer bound is in general not tight for exact-repair. Though a complete solution for the special case of $(4,3,3)$ is given, the rate region characterization problem under general parameters is still open. Readers may wonder if the procedure given in Section \ref{sec:CAP} can be used on the general problem, if it fundamentally alters the complexity order of the primal optimization problem. Unfortunately, although a few more cases with small $(n,k,d)$ values can be tackled this way, the complexity is still too high for larger parameter values, and the list of information inequalities involved in the proof is quite large. In fact, running only the set growth and symmetry determination procedures alone for each variable is of exponential order in the total number of random variables. As an on-going work, we are investigating whether low complexity procedures exist that can further take into account the symmetry. Through such a general study, we hope to discover more structure in the converse proof which may lead to the complete solution of general $(n,k,d)$ exact-repair regenerating codes.

%As an ongoing work, we are currently investigating the generalization of the results presented here for other $(n,k,d)$ parameters, and have obtained partial results on several more cases, which will be presented in a follow-up work.
%As an ongoing work, we are currently investigating the generalization of the result presented here for other $(n,k,d)$ parameters. 

%Though the special case of $(n,k,d,)=(4,3,3)$ is solved here, the general rate region characterization problem for exact-repair regenerating codes is still quite open. As an ongoing work, we are currently investigating the generalization of the result presented here for other $(n,k,d)$ parameters. 

%For an information theorist, an information theoretical proof using the fewest steps will be ideal, and for this purpose, a second LP problem is introduced with the objective of obtaining a sparse solution, which leads to an explicit algebraic proof.

\appendix

\begin{IEEEproof}[Proof of Proposition \ref{prop:symmetry}]
For any $(\bar{\alpha},\bar{\beta})$ that is $(4,3,3)$ exact-repair achievable, there exists for any $\epsilon>0$, an $(N,K_d,K)$ exact-repair regenerating code such that
\begin{align}
\bar{\alpha}+\epsilon\geq \frac{\log K_d}{\log N},\quad
\bar{\beta}+\epsilon\geq \frac{\log K}{\log N}. 
\label{eqn:conditiondefinition}
\end{align}
Let us for now fix a value $\epsilon>0$, and consider an $(N,K_d,K)$ exact-repair regenerating code satisfy the above conditions, which may or may not induce a symmetric entropy vector. Let the encoding and decoding functions be denoted as: $f^E_i(\cdot)$, $f^D_{A}(\cdot,\cdot,\cdot)$, $F^{E}_{i,j}(\cdot)$, $F^{D}_{i}(\cdot,\cdot,\cdot)$, as given in Definition \ref{def:NKKcode}. We shall show that it can be used to construct an $(N\rq{},K\rq{}_d,K\rq{})=(N^{24},K^{24}_d,K^{24})$ code that induces a symmetric distortion vector, which clearly satisfies (\ref{eqn:conditiondefinition}), and the proof will be completed by making $\epsilon$ arbitrarily small. 

%Recall the set permutation notation given after Definition \ref{def:permutation}, and define the following permuted functions
%\begin{align} 
%&f^E_i(\pi,\cdot) = f^E_{\pi(i)}(\cdot),\quad &f^D_{A}(\pi,\cdot)=f^D_{\pi{A}}(\cdot),\nonumber\\
%&F^{E}_{i,j}(\pi,\cdot)=F^{E}_{\pi(i),\pi(j)}(\cdot),\quad &F^{D}_{i}(\pi,\cdot)=F^{D}_{\pi(i)}(\cdot).
%\end{align}
Let the $24$ distinct permutations of $I_4$ be $\pi_1,\pi_2,\ldots,\pi_{24}$, and let their inverse function be $\pi^{-1}_1,\pi^{-1}_2,\ldots,\pi^{-1}_{24}$. The new encoding and decoding functions can be written as
\begin{align}
&\hat{f}^E_i\left(x^{(1)},x^{(2)},\ldots,x^{(24)}\right)= \prod_{k=1}^{24}f^E_{\pi^{-1}_k(i)}\left(x^{(k)}\right),\nonumber\\
&\hat{f}^D_{A}\left((x^{(1)}_1,x^{(1)}_2,x^{(1)}_3),\ldots,(x^{(24)}_1,x^{(24)}_2,x^{(24)}_3)\right)\nonumber\\
&\qquad\qquad\qquad\qquad\qquad=\prod_{k=1}^{24}f^D_{\pi_k^{-1}{A}}\left(x^{(k)}_1,x^{(k)}_2,x^{(k)}_3\right),\nonumber\\
&\hat{F}^{E}_{i,j}\left(x^{(1)},x^{(2)},\ldots,x^{(24)}\right)=\prod_{k=1}^{24}F^{E}_{\pi_k^{-1}(i),\pi^{-1}(j)}\left(x^{(k)}\right),\nonumber\\
&\hat{F}^{D}_{i}\left((x^{(1)}_1,x^{(1)}_2,x^{(1)}_3),\ldots,(x^{(24)}_1,x^{(24)}_2,x^{(24)}_3)\right)\nonumber\\
&\qquad\qquad\qquad\qquad=\prod_{k=1}^{24}F^{D}_{\pi_k^{-1}(i)}\left(x^{(k)}_1,x^{(k)}_2,x^{(k)}_3\right).
\end{align}
Because of the symmetry of the new encoding and decoding functions, it is clear by utilizing (\ref{WSdefinition}) that they indeed induce a symmetric entropy vector, according to Definition \ref{def:symmetricentropy}. The zero-error decoding and repair requirements are satisfied because the original code is able to accomplish them. The proof is thus complete.
\end{IEEEproof}

\section*{Acknowledgment}
The author wishes to thank Dr. Dahai Xu at AT\&T Labs-Research for introducing him to the Cplex optimization software package. 

\bibliographystyle{IEEEbib}

\end{document}